# Model-free Analysis of Scattering and Imaging Data with Escort-Weighted Shannon Entropy and Divergence Matrices

Jared Coles [1,2], Arthur R. C. McCray [1,3], Yue Li [1], Bryan T. Fichera [1], Yan Wu [4], Yiqing Hao [4], Daniel Phelan [1], Yue Cao [1], Raymond Osborn [1], C. Phatak [1], Stephan Rosenkranz [1], Yu Li [1]

[1] Materials Science Division, Argonne National Laboratory, Lemont, Illinois 60439, USA

[2] Department of Physics, Northern Illinois University, DeKalb, IL 60115, USA

[3] Department of Materials Science and Engineering, Stanford University, Stanford, CA, USA

[4] Neutron Scattering Division, Oak Ridge National Laboratory, Oak Ridge, Tennessee 37831, USA

We demonstrate a model-free data analysis framework that leverages escort-weighted Shannon Entropy and several divergence matrices to detect phase transitions in scattering and imaging datasets. By establishing a connection between physical entropy and informational entropy, this approach provides a sensitive method for identifying phase transitions without an explicit physical model or order parameter. We further show that pairwise divergence matrices, including Kullback-Leibler divergence, Jeffrey Divergence, Jensen-Shannon Divergence and antisymmetric Kullback-Leibler divergence, provide more comprehensive measures of statistical changes than scalar entropy alone. Our approach successfully detects the onset of both long- and short-range order in neutron and X-ray scattering data, as well as a non-trivial phase transition in magnetic skyrmion lattices observed through Lorentz-transition electron microscopy. These results establish a framework for automated, model-free analysis of experimental data with broad applications in materials science and condensed matter physics.

## 1. Introduction

Entropy stands as one of the most fundamental concepts in physics. It is intimately linked to the second law of thermodynamics and is central to our understanding of diverse phenomena, from frustrated magnetism [1,2] and black hole radiation [3] to emergent gravity [4] and information theory [5,6]. In thermodynamics, entropy provides a measure of disorder and is generally defined via the Gibbs formulation, $S_{Gibbs} = -k_B \sum p_i \ln(p_i)$, where $p_i$ is the probability of occupying a microstate $i$, and $k_B$ is the Boltzmann constant. Similarly in information theory, the Shannon entropy which possesses a similar mathematical form (defined in the next section), quantifies the average amount of information or uncertainty in a dataset. While thermodynamic entropy is often regarded as a physical quantity, Shannon entropy is typically viewed as "subjective", measuring the incompleteness of an observer's knowledge of a system's state. E.T. Jaynes, in one of his seminal works, proposed a fundamental conceptual connection between the two formulations of entropy [7,8], where the thermodynamic entropy is derived from Shannon entropy through principles of statistical inference. This connection was further developed by Landauer [9] and experimentally validated in several studies [10,11], which showed that erasing information necessarily leads to the dissipation of heat, revealing an intrinsic entropy cost to information processing. It is now well established that there exists an entire family of entropy measures, and some of them have attracted renewed interest for their usefulness in quantifying entanglement [12] and other physical phenomena across a range of scientific disciplines.

Here we focus on scattering and imaging experiments, where the image and volumetric data encode information about the physical state of a material. Traditional data analysis often focuses on identifying peaks in scattering spectra (e.g., as a function of energy or momentum transfer). In condensed matter physics, the intensities of peaks at ordering wave vectors can be used to track order parameters as functions of thermodynamic control parameters, such as temperature, pressure, and applied magnetic or electric fields [13-15]. However, this approach is only useful if the ordering wave vector can be identified; there is a risk of oversight if ordering occurs at unknown wave vectors that are unrecognized by experimenters or data analysis routines. This risk has been reduced by the introduction of large-area detectors, which dramatically expand reciprocal-space coverage [16] and enable repad collection of large datasets with many potential regions of interest. Nevertheless, the resulting increase in data volume creates significant challenges for rapid analysis, as conventional model-drive methods can be time-consuming and often require sophisticated techniques [17,18].

To address these challenges, we assume a hypothetical mapping between the physical state distribution and the distribution of intensities in measurement (data) space:

$$\Phi(x'_1, x'_2, \dots) = \mathcal{F}(P(x_1, x_2, \dots)),$$

where $\Phi(x'_1, x'_2, \dots)$ denotes a probability distribution in a high-dimensional data space with coordinates $x'_1, x'_2,\dots$, and $P(x_1, x_2, \dots)$ is the probability distribution over the system's phase space with physical degrees of freedom $x_1, x_2, \dots$. By connecting the thermodynamic entropy of the sample to the informational entropy of the measured datasets (Fig. 1) through this mapping, we develop an information-theory-inspired analysis framework that operates directly on measured data without requiring an explicit physical model. Based on the escort distribution and relative entropy (divergence), we demonstrate that this framework captures the onset of long- and short-range magnetic and structural order, as well as a non-trivial two-dimensional phase transition in magnetic skyrmion lattices. We present case studies spanning a range of data types, including neutron diffraction data on $Eu_3Sn_2S_7$, X-ray scattering on $Cd_2Re_2O_7$ [18] and Lorentz-transition electron microscopy (LTEM) images of magnetic skyrmions in $Fe_3GeTe_2$ [19]. We show that escort-weighted entropy and pairwise divergence matrices serve as effective and model-free indicators of potential phase changes and transitions. We emphasize that this development will not replace conventional physics-based analysis but instead provides a complementary method to rapidly identify candidate phase transitions. Our framework can be integrated into existing analysis pipelines and used as input for advanced machine-learning workflows [17]. These results highlight opportunities to leverage informational and statistical tools for automated, model-free analysis of large scientific datasets.

## 2. Theoretical and Methodological Framework

### 2.1 Escort Distribution and Artificial Temperature

In general, scattering or imaging datasets in one- or multi- dimensional spaces can be normalized into a probability distribution $P = \{p_i\}$, where $p_i$ represents the normalized intensity at pixel or voxel $i$, and satisfies the normalization condition $\sum_i p_i = 1$. Such datasets can be represented in various formats, such as detector pixels, wave vectors in $\boldsymbol{Q}$-space, positional indices in real space, or even time steps along a temporal axis. The Shannon entropy, quantifying the information content of a distribution, is defined by:

$$H(p) = -\sum_i p_i \log_2 p_i$$

In a typical diffraction experiment on a perfectly ordered crystal, the diffraction intensity distribution exhibits Bragg peaks that are ideally $\delta$-functions, with zero intensity elsewhere. The low thermodynamic entropy of a crystal with infinite long-range order is reflected in the minimal Shannon entropy of the corresponding diffraction dataset [Fig.1 (b)]. In contrast, if the atoms are completely random, with neither long- nor short-range order, the scattering pattern approaches a uniform distribution, and the information entropy is maximized.

While this connection is intuitive and not mandated by fundamental principles, it is reasonable to postulate that the transformation functional

$$\Phi(\boldsymbol{x}') = \mathcal{F}\big(P(\boldsymbol{x})\big)$$

is monotonic in the sense that it preserves the ordering of state probabilities. Specifically, if the physical distribution satisfies

$$p_1 \leq p_i \leq \cdots \leq p_j \leq \cdots \leq p_N,$$

where N is the total number of states (irrespective of symmetry or periodicity), then the transformed distribution preserves this ordering:

$$q_1 \leq q_i \leq \cdots \leq q_j \leq \cdots \leq q_N.$$

To formalize this idea, we adopt the escort distribution transformation [20, 21], defined by

$$q_i = \frac{p_i^n}{\sum_{k=1}^N p_k^n}$$

where $p_i$ represents the probability associated with pixel/voxel $i$ in $P(x)$, $q_i$ is the corresponding probability in $\Phi(x')$, and $n = \frac{1}{T_a}$. Here $T_a$ acts as an effective "artificial temperature" that controls the relative weight of each state (pixel or voxel). Apparently, this mapping is monotonic for $n > 0$, and the original distribution is recovered when $n = 1$ (equivalently, $T_a = 1$).

**2.2 Relative Entropy and Divergence Matrices**

While Shannon entropy characterizes the overall content of information in a dataset, changes across a phase transition as a function of external parameters can be subtle. For instance, in a neutron diffraction study of temperature-dependent magnetic order, the Shannon entropy might be dominated by the nearly constant intensity of structural Bragg peaks across the transition. To increase sensitivity to such changes, we introduce relative entropy, mirroring the Kullback-Leibler divergence (KLD) [22] in information theory:

$$D_{\mathrm{KL}}(P \parallel Q) = \sum_i p_i \log\left(\frac{p_i}{q_i}\right)$$

Here, $D_{\mathrm{KL}}(P \parallel Q)$ denotes the divergence between two probability distributions $P = \{p_i\}$ and $Q = \{q_i\}$ measured under two different conditions (e.g., temperatures or fields); in our divergence matrix, it corresponds to the matrix element comparing those two conditions. In general, KLD measures statistical difference between two probability distributions and quantifies the information loss when approximating $P$ with $Q$. Notably, this measure is asymmetric. In general, $D_{\mathrm{KL}}(P \parallel Q) \neq D_{\mathrm{KL}}(Q \parallel P)$. In computer science, symmetric variants are more broadly used, including the Jeffrey divergence (JD) [23]:

$$D_{\mathrm{J}}(P \parallel Q) = \frac{1}{2} D_{\mathrm{KL}}(P \parallel Q) + \frac{1}{2} D_{\mathrm{KL}}(Q \parallel P),$$

and the Jensen-Shannon divergence (JSD) [24]:

$$D_{\mathrm{JS}}(P \parallel Q) = \frac{1}{2} D_{\mathrm{KL}}(P \parallel M) + \frac{1}{2} D_{\mathrm{KL}}(Q \parallel M),$$

where $M = \frac{1}{2}(P + Q)$.

A key advantage of JSD is that it is bounded: $0 \leq D_{\mathrm{JS}}(P \parallel Q) \leq \log 2$, which makes it widely used in bioinformatics, social sciences, and deep learning [25, 26].

In addition to these symmetric forms, we introduce the antisymmetric KLD (a-KLD):

$$D_{\mathrm{aKL}}(P \parallel Q) = D_{\mathrm{KL}}(P \parallel Q) - D_{\mathrm{KL}}(Q \parallel P),$$

which measures directional changes of entropy between two datasets. We note that these asymmetric changes should not be directly interpreted as thermodynamic irreversibility and decide to leave further discussion of its physical interpretation for future work. As demonstrated below, these divergence matrices exhibit block-like structures whose boundaries separate statistically similar datasets from dissimilar ones, serving as visual indicators of phase transitions.

## 3. Experimental methods

We selected three full sets of representative temperature/field-dependent data from neutron scattering, X-ray scattering and L-TEM images for demonstration. Neutron diffraction data on a single crystal of $Eu_3Sn_2S_7$ was collected on the HB-3A (DEMAND) beamline at Oak Ridge National Laboratory. Both temperature and field dependencies were measured. Using a two-dimensional area detector, typical low-temperature data contain a single magnetic Bragg peak at wave vector $Q$ = (-0.5,1.5,0) r.l.u. (reciprocal lattice units) and surrounding background, as shown in Fig.2(a). This material exhibits an interesting magnetic Cairo lattice [27, 28] and undergoes two magnetic transitions at 6K and 4K at zero field, plus a series of metamagnetic transitions under applied magnetic fields. While the magnetic structure has not been previously reported, we leave detailed discussion of physics for a separate paper and focus here on identifying magnetic phase transitions from the dataset.

The second dataset comprises temperature-dependent X-ray diffraction on $Cd_2Re_2O_7$, which exhibits a series of intriguing structural transitions still under active debate [29-33]. While X-ray scattering data were published previously [18], we focus here on testing our entropy analysis on this representative diffuse scattering dataset containing thousands of Bragg peaks and short-range diffuse scattering, in contrast with our first example of neutron diffraction dataset that contained a single Bragg peak. The experiment was conducted at Sector 6-ID-D at the Advanced Photon Source using monochromatic X-rays with energy of 87 keV. Detector images were collected on a fast area detector (Pilatus 2M CdTe) at a frame rate of 10 Hz while the sample is continuously rotated over $360°$ at a speed of $1°/s$. After processing through our data reduction workflow in NeXpy, we conducted the entropy analysis on the resulting Q-space data.

In the third case, we extend beyond the scattering data and show the robustness of our framework on real-space L-TEM images of $Fe_3GeTe_2$. Previous L-TEM results reveal thermally hysteretic behavior in $Fe_3GeTe_2$, dependent on the strength of the out-of-plane magnetic field [19]. A 2D phase transition associated with the Néel skyrmion lattice occurs in the 150K – 200K temperature range for both field-cooled and field-heating protocols. Here we focus on the same published dataset collected during field heating that exhibits a relatively sharp change in skyrmion order under an applied field of 500 G. In the LTEM experiment, skyrmion lattice order, quantified by the local orientational order parameter, indicates a phase transition near 200K [19]. LTEM images were taken in Lorentz mode on a JEOL 2100F TEM operating at an accelerating voltage of 200 kV. Liquid nitrogen holder was used to cool the sample.

## 4. Demonstration of entropy analysis in experimental datasets

### 4.1 Temperature dependent neutron diffraction in $Eu_3Sn_2S_7$

Figure 2 (a) shows a raw detector image with a distinct magnetic Bragg peak at $\boldsymbol{Q} = (-0.5,1.5,0)$ r.l.u. at $T = 1.7\ K$. The peak intensity decreases with increasing temperature and is absent at $T = 10\ K$. The strong, localized intensity against near-zero background visually indicates a magnetically ordered state at low temperature. As shown in Fig. 2 (b), the peak intensity vanishes above $\sim 6\,K$, consistent with an antiferromagnetic transition near this temperature. The Shannon entropy computed from these images increases with temperature and saturates above 6K (Fig. 2(b)). This correspondence suggests that the Shannon entropy computed from raw data images captures the evolution of magnetic order in this material, with lower entropy corresponding to more ordered states.

Fig. 2(c) displays the computed pairwise KLD matrix, which exhibits a square block below 6 K corresponding to the magnetically ordered phase. The low divergence values within this block reflect the statistical similarity among low-temperature states, distinct from high-temperature disordered phase. Similar block-like structures also appear in the symmetric JD and JSD matrices, as well as in the a-KLD matrix [Fig. 2 (d)-(f)]. Notably, an additional transition is marginally visible near 4K in Fig. 2(e). This feature is supported by magnetic susceptibility measurements and neutron diffraction at another wave vector, although it is not directly visible in temperature-dependent peak intensity in Fig. 2(b). Overall, these results suggest that our statistical analysis of raw neutron scattering data can identify potential phase transitions without requiring prior knowledge of the ordering wave vector or explicit model.

### 4.2 Field-induced transition in $Eu_3Sn_2S_7$

Since the analysis above does not explicitly require temperature as input, we can extend this approach to detect magnetic field-dependent phase transitions. Figure 3 (a) shows the field-dependence ($\mu_0 H$) of the magnetic peak intensity at $\boldsymbol{Q} = (-0.5, 1.5, 0)$. Three prominent features are observed: two sharp intensity drops signaling field-induced metamagnetic transitions at $\mu_0 H = 0.5$ T and 2 T, and a change in slope above 3T consistent with magnetization saturation. Figure 3(b) shows the Shannon entropy as a function of field; however, it is noisy, probably reflecting the contribution of background. In contrast, the escort-weighted entropy with exponent n=2, plotted in the same figure, effectively reduces the weight of background (discussed in section. 5) and clearly reveals the two field-induced transitions around 0.5 T and 2 T. Figure 3 (c)-(f) displays the four divergence matrices computed based on the escort distribution ($n = 2$). Their color maps exhibit similar block-like structures whose boundaries align with the transitions identified from the peak intensity in Figure 3 (a). As we have shown, introducing the escort distribution into the calculation of entropy and divergences significantly enhances the sensitivity and robustness of entropy analysis for detecting subtle phase transitions.

### 4.3 Detecting diffuse scattering and multiple transitions in $Cd_2Re_2O_7$

Owing to the brightness of synchrotron sources and development of pixelated detectors, it is now efficient to collect large (> 1 GB) datasets with X-ray scattering. To examine the applicability of the entropy analysis to large datasets, we applied it to our existing X-ray diffuse scattering data on $Cd_2Re_2O_7$, which exhibits intriguing phase transitions and diffuse scattering. Figure 4 (a) presents a slice of scattering data in $Q$-space at 150 K, with both Bragg peaks and diffuse scattering clearly observed. Figure 4 (b) shows the computed Shannon entropy (n=1) and the escort-weighted entropy (n=0.5). While the two curves exhibit completely different behaviors, they both show features across the phase transitions near 200 K and 100 K. We further computed all four divergence matrices using different values of $n$ ranging from 0.125 to 1, and they all consistently exhibit the same pattern with block-like structures as shown in Figure 4 (c)-(f), indicating a suitable range for the choice of n. In Figure 4 (e), the a-KLD shows five blocks along the diagonal direction, separated by four transitions at approximately 100 K, 130 K, 200 K, and 250 K. While signatures of these transitions are marginally visible in the other three divergences in Fig. 4 (c), (d) and (f), it appears that the a-KLD is more sensitive to subtle phase transitions, which was also seen in the neutron diffraction studies in Fig.2 and Fig. 3. We note that the strong deviation in the JSD between 90K and 100K in Fig. 4 (f) originates primarily from switching the cryogen from helium to nitrogen, indicating that these matrices are also sensitive to background variations. Nevertheless, additional analysis focused on smaller areas of interest provides evidence for a phase transition in this temperature range.

Previous studies have suggested that $Cd_2Re_2O_7$ exhibits a second-order phase transition at $T_{s1} = 200\ K$ from the cubic pyrochlore $Fd\bar{3}m$ structure to an intermediate phase related to the $I\bar{4}m2$ component of $E_u$ symmetry [29], a first-order transition at $T_{s2} = 113\ K$ associated with the $I4_122$ space group, and a possible third transition [29] into orthorhombic symmetry around $T_{s3} = 90\ K$, accompanied by the collapse of diffuse scattering intensities [18]. Although there exists a slight difference in phase transition temperatures from our analysis approach, we can attribute the three low-temperature transitions around 100 K, 130 K, and 200 K from our entropy analysis to the three reported transitions in the literature. Furthermore, we find the fourth transition at 250 K coincides with the rise of diffuse scattering near $(0\bar{4}6)$ and (115) peaks reported in the literature [18], likely related to the soft mode above $T_{s1}$. These results clearly demonstrate the advantage of our entropy analysis approach in identifying complex phase transitions in large datasets without requiring prior structural models.

### 4.4 Detection of phase change/transition of magnetic skyrmions in Fe$_3$GeTe$_2$

In the previous scattering experiments, the intensity at each pixel or voxel can be directly related to a specific q-state of the system via Fourier transformation. However, this connection is not as straightforward when processing real-space imaging datasets. In the following case, we demonstrate that the same entropy analysis technique can be applied to a broad range of data types beyond reciprocal-space measurements. Here, we use our previously published L-TEM datasets of Fe$_3$GeTe$_2$ which was found to host a transition in its 2-dimensional magnetic skyrmion lattice [19]. Figure 5 (a) displays a representative L-TEM image of a Néel skyrmion lattice at 100 K. Previous statistical analysis, with machine-learning-assisted image processing, indicated a transition from a disordered skyrmion lattice to a more ordered phase as the temperature approaches 200 K [19]. Figure 5 (b) presents both the Shannon ($n = 1$) and escort-weighted ($n = 2$) entropies, showing similar increases upon field-heating. These entropy changes coincide with the emergence of orientational partial ordering of magnetic skyrmions. Figure 5 (c)-(f) show the computed divergence matrices based on the escort distribution ($n = 2$). The color maps reveal a clear square block below ~180 K, with a crossover between 180 K and 210 K. The boundary between low-temperature and high-temperature states, however, is less sharp than those observed in Figures 2-4, likely due to the intrinsic 2D nature of the transition. Our published work suggests that this transition is associated with both long-range translational effects and local orientational order. While our entropy analysis does not directly probe the local orientational order parameter, our results in Fig. 5 might reflect the statistical variation associated with skyrmion sizes and/or densities. Nevertheless, the close correspondence between the material's phase evolution and the divergence matrices in Figure 5 (c)-(f) demonstrates that our entropy analysis framework robustly applies to real-space imaging datasets. Beyond this example, we also conducted additional tests on time-resolved X-ray photon correlation spectroscopy (XPCS) and electron scattering images, which further confirm the broad applicability of our analysis framework regardless of the data types and experimental probes.

## 5 Discussion

Having demonstrated the practical advantages of this entropy analysis framework, we now discuss its implications. In principle, experimental observation can be considered as an information transfer process, wherein the vast degrees of freedom in a physical system are reduced to finite datasets via an unknown transformation (Fig.1) depending on the experimental methods and parameters, such as instrument geometry, coherence loss, limited elemental contrast, finite detector coverage, etc. Therefore, only a fraction of the total information can be captured in a single experiment, which may not suffice to uniquely determine the ground truth. In fact, comparing measured observables from complementary experimental approaches is necessary and frequently used in discovering hidden phases in the past. For instance, in water ice [34,35], it is well-known that thermodynamic measurements fail to capture the partially frozen degrees of freedom according to the so-called "ice rule", resulting in the discovery of residual entropy when compared to spectroscopic measurements. A similar practice was also conducted by A. Ramirez in the discovery of zero-point entropy in spin ice [2, 36, 37], defined as the difference between the measured entropy from specific heat data and that of an assumed high-temperature paramagnetic phase.

Although there exists considerable ambiguity in interpreting the computed entropy, some intuitive understanding can be obtained for typical scattering experiments, which is essentially a Fourier transformation of atomic positions or electron density distribution, with the weight of each pixel (voxel) corresponding to the probability of a specific $Q$-state of the system. Due to the incomplete coverage of the reciprocal space, however, the measured probability distribution contains partial information, representing a conditional probability distribution [8] rather than the true distribution. Our introduction of the escort distribution essentially establishes an approximate connection between the conditional probability distribution and the ground truth. While such an approximation cannot be strictly proved, we emphasize that the advantage of our entropy analysis framework lies in its practical benefits, rather than mathematical rigor.

We can gain further insights into our entropy analysis through analogy with thermodynamic entropy. First, the divergence matrices quantify statistical changes in entropy between different macroscopic states separated by a small change of an external parameter (e.g., $\Delta T$) and are therefore analogous to a specific heat measurement based on small temperature increments. By taking a line cut slightly off the diagonal of a divergence matrix, such as $D_{KL}(T, T + \Delta T)$ from the KLD matrix, one can obtain a curve analogous to specific heat, where a peaks or shoulder (manifested as a corner of the block structure) indicate a phase transition in the data space.

Second, the introduction of the escort distribution with a tunable exponent $n$, or $T_a$, effectively controls the "temperature" of the measured datasets. For instance, an escort distribution with $n > 1$ (equivalently $T_a < 1$) amplifies high-valued, more singular regions of the original distribution, analogous to cooling a system and concentrating population in low-energy states. Conversely, for $n < 1$, it enhances less-populated regions, resembling heating. To illustrate how $T_a$ modulates the estimated entropy, we present a phase diagram of the escort-weighted entropy with varied $T_a$ in Figure 6, using the field-dependent datasets of $Eu_3Sn_2S_7$. Because the calculated entropy can vary by several orders of magnitude, we visualize the results by normalizing the minimum and maximum of each field-dependent entropy curve (horizontal line in Fig. 6) to 0 and 1 respectively. The artificial temperature $T_a$ is plotted on a logarithmic scale, with $\log_2 T_a = 0$ corresponding to the original Shannon entropy.

In Fig.6, we find that the escorted-weighted entropy is noisy at high artificial temperatures, but some visible transitions appear near the field of ~2 T. As $T_a$ is reduced, the computed escort-weighted entropy enters a critical region in which its field dependence exhibits well-structured features with a visibly high signal-to-noise ratio. As $T_a$ decreases further, this region terminates at a lower bound, beyond which the entropy curve becomes noisy again. In the limit $T_a \to 0$, the escort distribution is expected to condense toward a "ground state" dominated by one or a few pixels/voxels, while the rest carry zero weight. Within the critical region of $T_a$, we found that the divergence matrices consistently produce block-like features, indicating that a phase diagram like Fig. 6 can be used as a guide for selecting a suitable value for $T_a$. Interestingly, in all the datasets we have tested (Figure 2-5), we found that the optimal value of $T_a$ falls into the region with $\log_2 T_a$ close to 0. This means the raw probability distribution without "temperature tuning" approximately represents the real physical system. Therefore, the value of $T_a$ might also be useful as a rough indicator of the "efficiency" of a measurement. Considering that each neutron or X-ray instrument is specially designed for certain types of measurements, we suspect the choice of $T_a$, although empirical, may remain within a reasonably small range for a particular instrument configuration. This opens the possibility of on-the-fly implementation of entropy analysis across a broad range of scattering and imaging experiments.

## 6 Conclusion

In conclusion, we have developed an entropy-based analysis framework that connects the thermodynamic entropy of materials and the informational entropy in experimental datasets. By transforming measured data into an escort distribution, we can associate changes in Shannon entropy of the datasets with the formation of various physical orders in materials, including long- and short-range magnetic and structural orders as well as non-trivial orientational order in topological spin textures. We introduced four divergence matrices (KLD, JD, JSD and a-KLD) that collectively provide comprehensive measures of potential phase transitions without explicit models or prior knowledge. We demonstrated that this framework is broadly applicable to diverse datasets and experimental probes, including neutron/X-ray scattering, and real-space imaging. As data collection rates in modern experiments continue to increase with technical advances such as large area detectors and higher flux, model-free informatics offers a timely and efficient complement to traditional model-dependent analysis. Future instrumental developments, such as on-the-fly entropy analysis operating directly in raw data space and integration into advanced machine learning workflow, may provide new insights into increasingly complex experimental datasets.

## Acknowledgments

Y. L. and S. R. thank Jacob Ruff for insightful discussions regarding image compression and entropy estimation. This work was supported by the U.S. Department of Energy, Office of Science, Basic Energy Sciences, Materials Sciences and Engineering Division. Y.C. and B.F. further thank the U.S. Department of Energy, Office of Science for the Early Career Research Program for simulating the atomic disorder in ideal systems. This research used resources at the Advanced Photon Source, a U.S. DOE Office of Science User Facility operated by Argonne National Laboratory. A portion of this research used resources at the High Flux Isotope Reactor, a DOE Office of Science User Facility operated by the Oak Ridge National Laboratory. The beam time was allocated to HB-3A on proposal number IPTS-32168 and IPTS-33344.

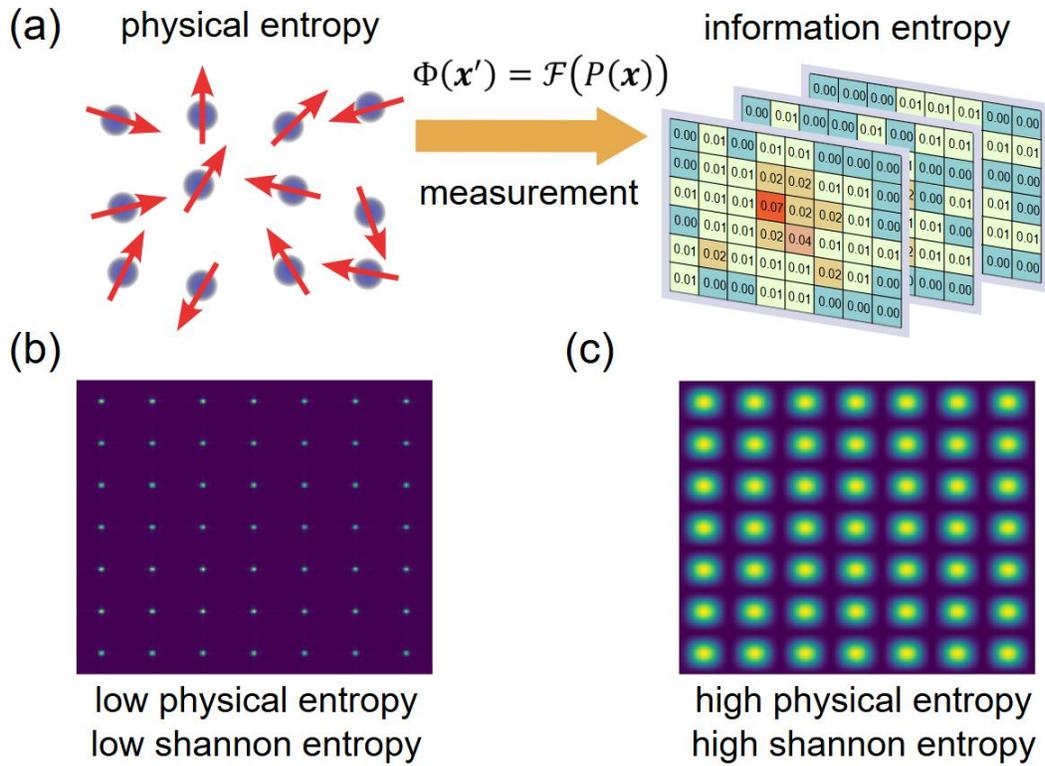

**Figure 1** (a) Schematic illustration of the relationship between physical entropy and information entropy. (b) Diffraction intensity from perfectly ordered materials where low physical entropy corresponding to low Shannon entropy. (c) Diffraction intensity from disordered materials, where high physical entropy is reflected by Shannon entropy in the data.

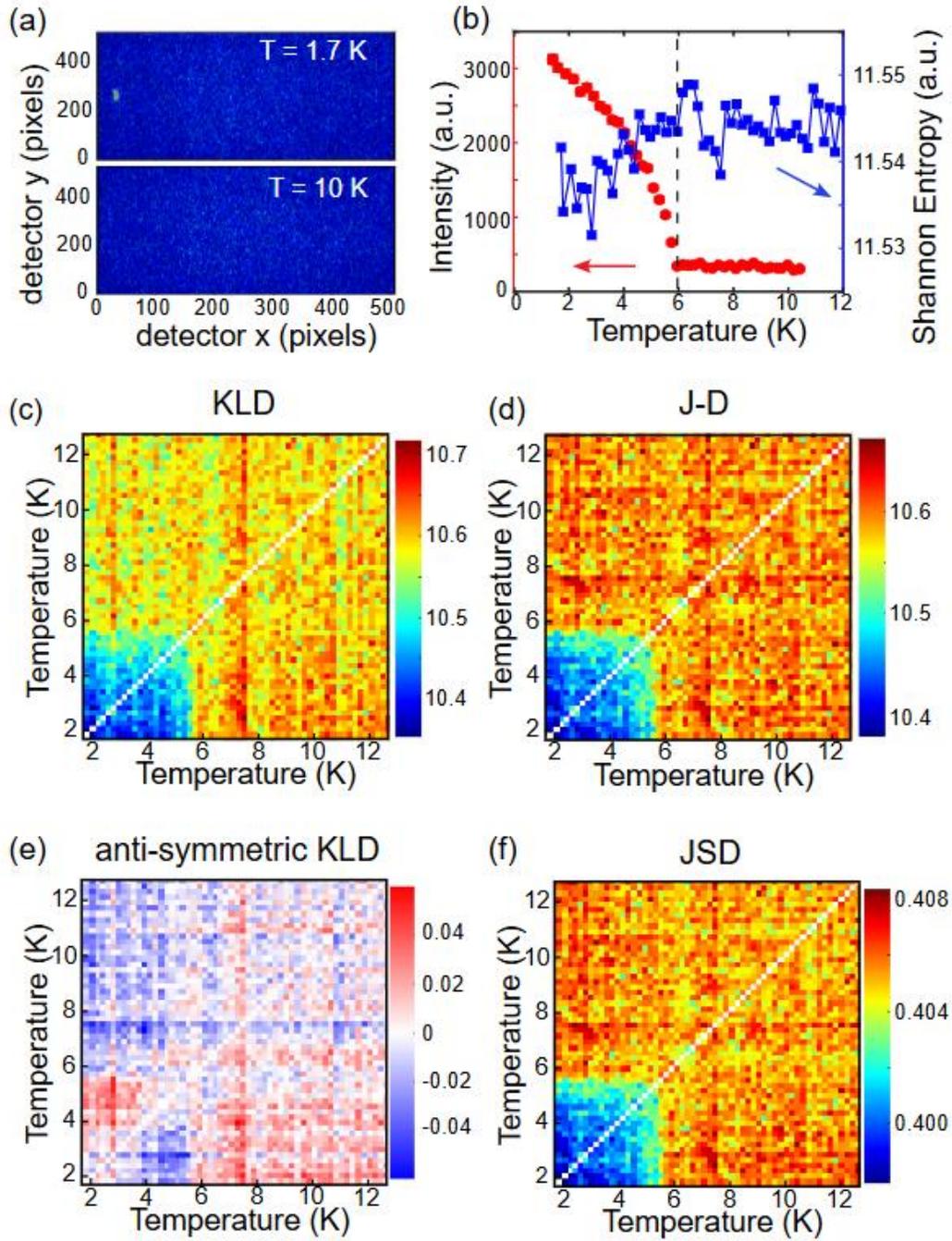

**Figure 2** (a) Detector image of the magnetic Bragg peak in $Eu_3Sn_2S_7$ at q = (-0.5,1.5,0). (b) Temperature dependence of the peak intensity (red) and the Shannon entropy (blue). (c)-(f), KLD, JD, a-KLD and JSD matrices computed from the detector images at a series of temperatures. Blocks with similar colors in the divergence matrices indicate statistically similar states, and their boundaries correspond to the phase transition at 6 K.

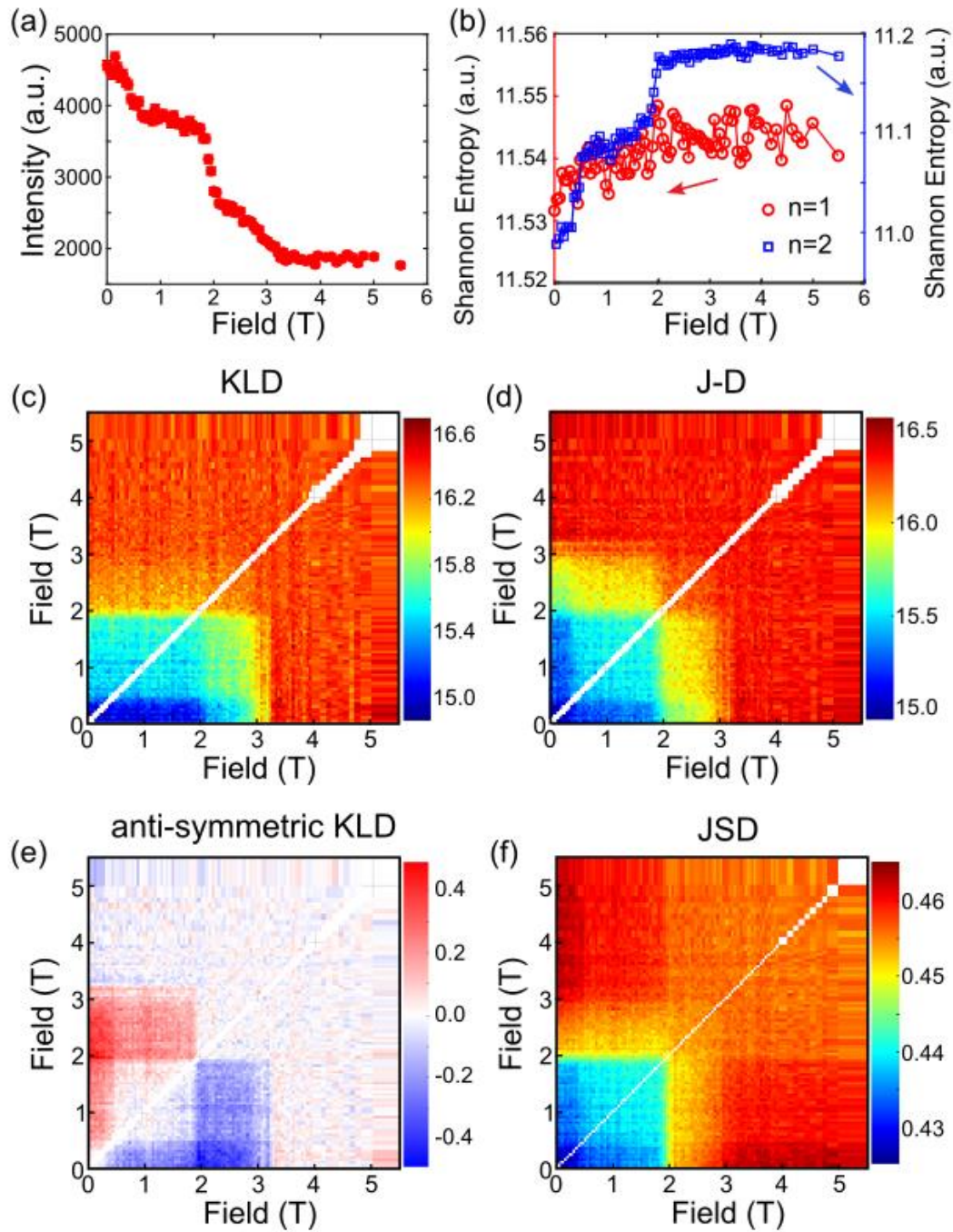

**Figure 3** (a) Magnetic field dependence of the magnetic Bragg peak intensity at Q = (-0.5,1.5,0). (b) Shannon entropy (n=1) and escorted-weighted entropy (n=2) as functions of magnetic field. (c)-(f) Divergence matrices highlighting a series of field-induced transitions at 0.5T, 2T, and 3T.

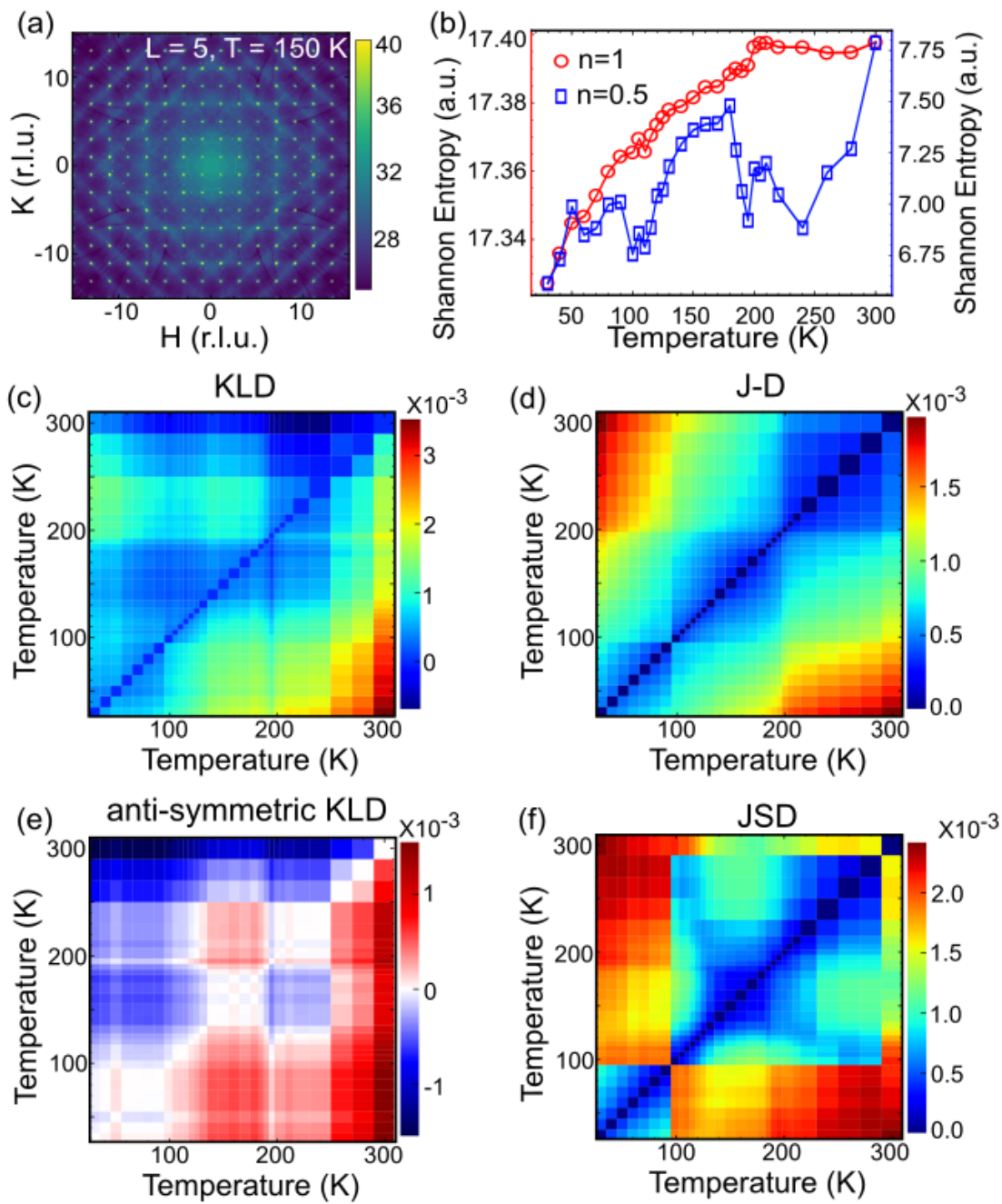

**Figure 4** (a) A [H,K,0] slice of the X-ray scattering data for $Cd_2Re_2O_7$ measured at 150 K. (b) Shannon entropy (n=1) and escort-weighted entropy (n=0.5) as functions of temperature. (c)-(f) Divergence matrices revealing transitions near 100K, 130K, 200K, and 250 K.

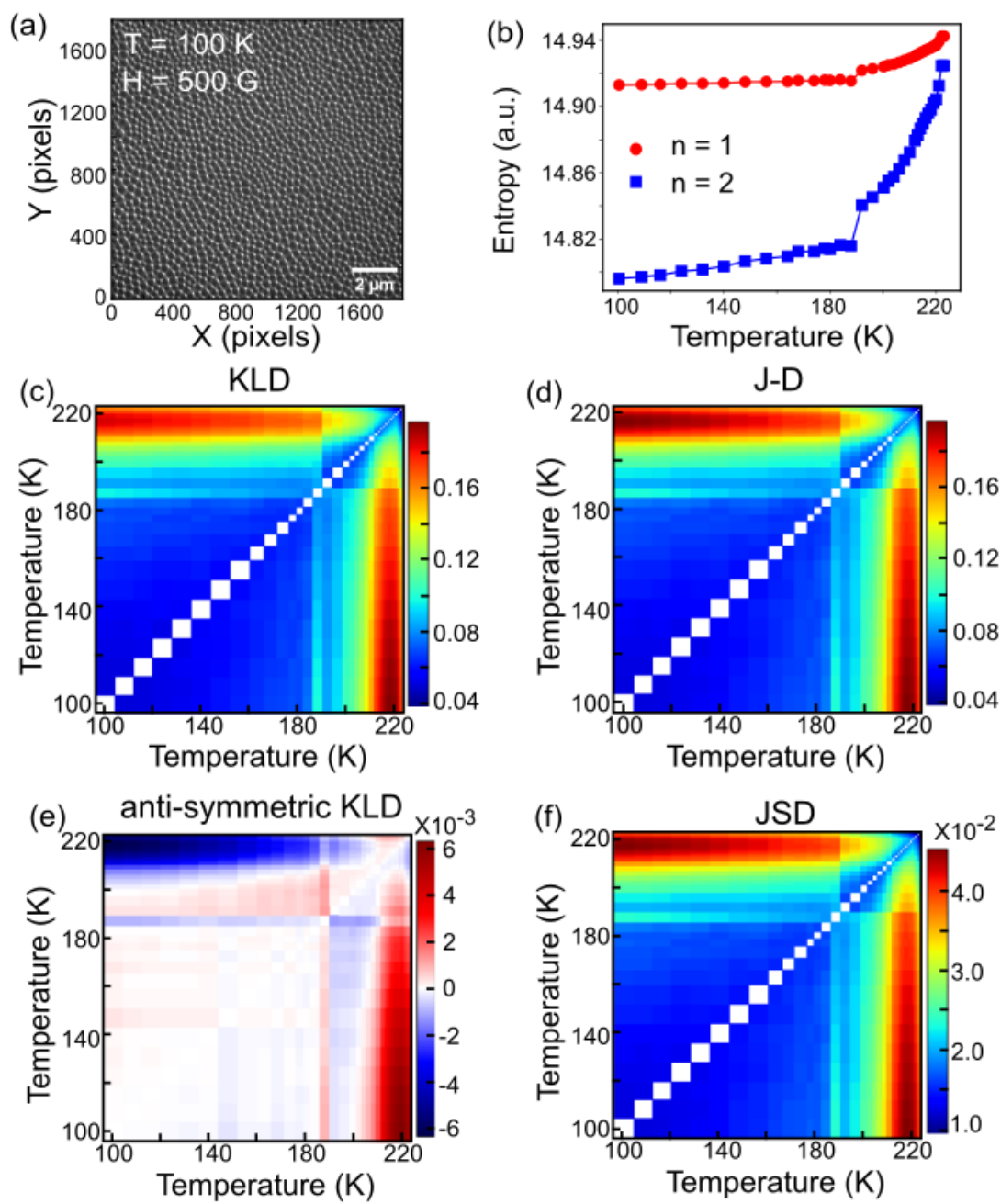

**Figure 5** (a) Lorentz-TEM image of magnetic skyrmions in $Fe_3GeTe_2$ at T= 100 K. (b) Shannon entropy (n=1) and escort-weighted entropy (n=2) as functions of temperature. (c)-(f) Divergence matrices highlighting the restoration of orientational order in the skyrmion lattice around 200K.

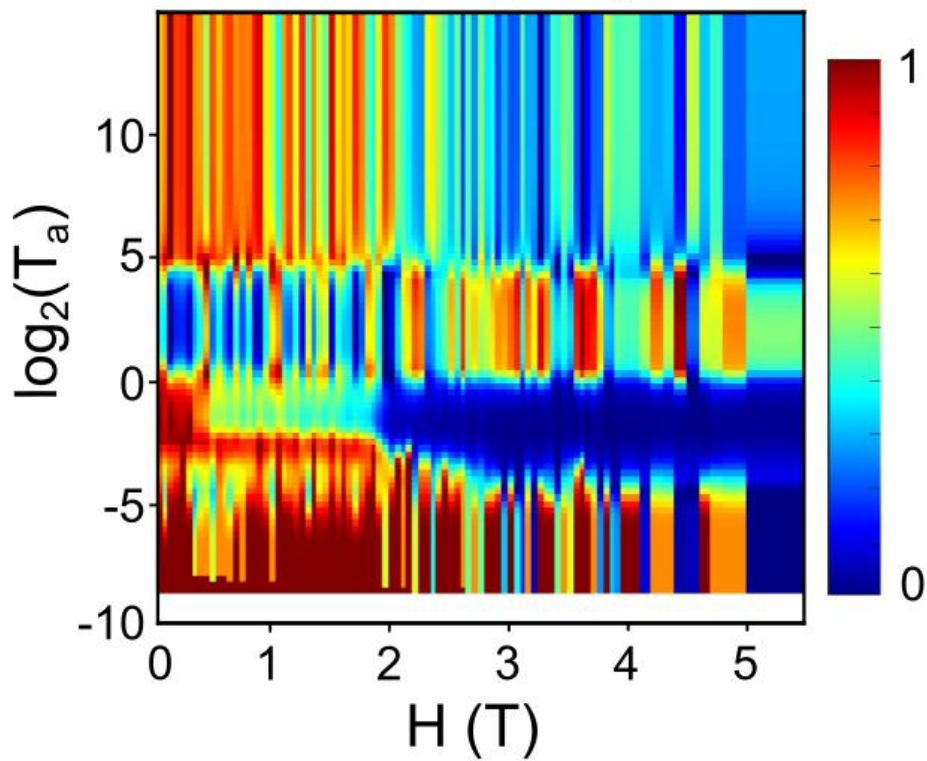

**Figure 6** Diagram of escort-weighted entropy as a function of $T_a$ from the field-dependent $Eu_3Sn_2S_7$ dataset. The horizontal axis denotes magnetic field, and the vertical axis is the artificial temperature $T_a$, shown on a logarithmic scale. Each row is normalized between its minimum and maximum to emphasize the relative variation within that row.